\documentstyle[aps,prl,twocolumn,epsf,psfig]{revtex} 
\begin{document}
\flushbottom
\draft
\twocolumn[
\hsize\textwidth\columnwidth\hsize\csname@twocolumnfalse\endcsname
\title{Multiexciton molecules in the hexaborides}
\author{T. A. Gloor, M. E. Zhitomirsky and T. M. Rice} 
\address{Institut f\"ur Theoretische Physik, ETH-H\"onggerberg, 
CH-8093 Zurich, Switzerland}
\date{\today} 
\maketitle 
\begin{abstract}
\hspace*{2mm}
We investigate multiexciton bound states in a semiconducting
phase of divalent hexaborides. Due to three degenerate valleys 
in both the conduction and valence bands the binding energy of 
a 6-exciton molecule is greatly enhanced by the shell effect. 
The ground state energies of multiexciton molecules are calculated 
using the density functional formalism. We also show that charged
impurities stabilize multiexciton complexes
leading to condensation of localized excitons. These
complexes can act as nucleation centers of local moments.
\end{abstract}
\pacs{PACS numbers:
                   71.35.--y, 
                   71.35.Lk, 
                   77.84.Bw  
}
\vspace{5mm} 
]

\narrowtext

\section{Introduction}
The recent discovery of unusual high-temperature weak ferromagnetism 
in lightly doped divalent hexaborides, CaB$_6$, SrB$_6$, BaB$_6$, 
\cite{Young99} has raised the need to understand the properties of these 
compounds in a broader perspective. The unique features of 
the novel type of ferromagnetism in Ca$_{1-x}$La$_x$B$_6$ are 
(i) there are no partially filled $d$- or $f$-orbitals, 
(ii) magnetism appears only at finite doping $0<x<2$\%, and 
(iii) tiny magnetic moments ($\sim 0.1\mu_B$ per doped electron) are 
very robust and develop at temperatures as high as 600--1000~K. 
All these features are consistently explained by ferromagnetism 
of a doped excitonic insulator \cite{Mike99a}. Thus, not only is
the magnetic phase quite unusual, but also the undoped stoichiometric 
hexaborides may also exhibit a novel type of ground state --- 
a condensate of bound electron-hole pairs or excitons. 
The theory of excitonic insulators has been developed long time ago,
see \cite{Halperin68}. However, divalent hexaborides seem 
to be the first experimental realization of an excitonic insulator 
made possible by their unique band structure. Band structure 
calculations \cite{Hasegawa79,Massida97} predict a small overlap of 
valence and conduction bands at the 3 symmetry related
nonequivalent $X$ points in the 
cubic Brillouin zone. There is a certain ambiguity both from 
theoretical \cite{Hasegawa79,Mike99a} and experimental 
\cite{Degiorgi,Terashima} points of view regarding whether 
the stoichiometric divalent hexaborides have a small band overlap or 
a small band gap. Precise treatment of correlation effects
in electron-hole plasma \cite{Mike99b} beyond the standard
approximations of band structure calculations
predicts a first order transition  from a semiconductor
to a semimetal under pressure. Therefore, stoichiometric hexaborides
may be on a semiconducting side of the phase diagram close to
the first order metal-insulator transition.

Motivated by the unique band structure with equal number of 
degenerate valleys in conduction and valence bands with different 
symmetries, we investigate here the formation of multiexciton 
molecules and complexes in these materials, assuming that there is a 
finite band gap between conduction and valence bands. 
In the effective mass 
approximation the valley index appears as an extra quantum number. 
By analogy with nuclei we expect a shell 
structure of the single particle orbits where six electrons and six 
holes with different spin and valley quantum numbers
can occupy the lowest shell. Hence, the 6-exciton molecule is 
expected to be the most stable configuration. 

Binding energy calculations of several excitons have been reported 
in the late seventies for multiexciton complexes bound to impurities 
in Si and Ge \cite{Kirczenow77,Dean76,Insepov77,Wunsche78b}. It was 
recognized that Hartree-Fock calculations do not work for multiexciton 
complexes because the main contribution to binding energy comes from 
correlation effects and that the density functional formalism provides 
a useful tool to include such correlation effects.
In this paper we consider both multiexciton molecules and multiexciton
complexes bound to impurities and calculate their binding 
energies within the framework of the density functional theory. 
The Kohn-Sham equations are single particle Schr\"odinger equations 
for electrons and holes moving in a self-consistent potential. These 
equations connect the density 
functional formalism to the shell model of multiexciton complexes and 
molecules \cite{Kirczenow77}.

The binding energies of exciton molecules provide information on
whether a Bose gas of multiexciton molecules is stabilized between 
semiconducting and semimetallic phases when the band gap is varied from 
positive to negative values. In the previous study by two of us 
\cite{Mike99b} an intermediate phase of a free exciton gas stabilized
by intervalley scattering processes was considered. Binding of several 
excitons is an alternative mechanism for appearance of an intermediate 
phase between a semiconductor and a semimetal. The binding energy per 
electron-hole pair in the molecule has to be compared to the minimal 
ground state energy of the electron-hole liquid, which according to
Ref.~\cite{Mike99b} is $E_{\rm g.s.}^{\rm min} = -1.55E_x$ in units of 
the excitonic Rydberg $E_x$. As another application of our results we 
suggest that the formation of multiexciton 
complexes near donor impurities 
(e.g., La-substitutions) can be a source 
of local magnetic moments, which 
appear in a semiconducting state. Such moments could explain the
unusual NMR 
relaxation rate measurements in the hexaborides \cite{Gavilano}.

\section{Method}
\label{section:Model} 

We use the effective mass approximation
considering electrons near minima in the conduction band and holes
near the maxima of the valence band as oppositely charged 
quasiparticles with anisotropic effective masses which interact 
via a screened Coulomb potential $V(r) = e_1e_2/\epsilon r$, 
$\epsilon$ being the static dielectric constant. 
For CaB$_6$ the principal values of the effective mass
tensor in units of the bare electron mass are given by 
$m_e^\parallel=0.504$, $m_e^\perp =0.212$ (conduction band) and 
$m_h^\parallel = 2.17$,  $m_h^\perp = 0.206$ (valence band).
The model Hamiltonian is
\begin{eqnarray}
 \hat{H} & = & \hat{T} + \hat{U} \ ,\label{eqn:modham} \\
 \hat{T} & = & \sum_\lambda\int d{\bf r}\;
 \Psi^\dag_\lambda({\bf r})
 K_\lambda\Psi_\lambda({\bf r}) \ , \nonumber \\
 \hat{U} & = & \frac{1}{2}\sum_{\lambda,\lambda'} \int d{\bf r} 
 d{\bf r'}\;\Psi^\dag_\lambda({\bf r}) \Psi^\dag_{\lambda'}({\bf r'})
 V_{\lambda,\lambda'}\Psi_{\lambda'}({\bf r'})\Psi_\lambda({\bf r})\ ,  
 \nonumber  \\
  & & V_{\lambda,\lambda'}=\frac{q_\lambda q_{\lambda'}}
 {\epsilon|{\bf r}-{\bf r}'|} \ , \nonumber
\end{eqnarray}
where $\lambda=(\kappa,\nu,\sigma)$ with $\kappa = e,h$; $\nu= 1,2,3$ 
is the valley index; $\sigma=\pm$ is the spin index;
and $q_\lambda$ is $+e$ for positively charged holes and $-e$ for
negatively charged electrons.  The kinetic energy operator 
$K_\lambda$ for one valley is 
\begin{equation}
\label{eqn:ekinani}
K_{(\kappa,1,\sigma)} = -\frac{\hbar^2}{2}\left[
\frac{1}{m_{\kappa}^\perp}\left(\frac{\partial^2}{\partial y^2}
  +\frac{\partial^2}{\partial z^2}\right)+\frac{1}{m_\kappa^\parallel}
\frac{\partial^2}{\partial x^2}\right].
\end{equation}
The expressions for the valleys $2$ and $3$ are obtained by cyclic
permutations of $x$, $y$ and $z$. We investigate the ground state 
energy of this Hamiltonian for a given number of electrons and holes. 
To treat this problem  we use the density functional formalism which 
is presented in the next section. In the following all equations are 
written in dimensionless variables. A natural set of units is the 
excitonic Rydberg for the energy scale $E_x = \mu e^4/2\hbar^2\epsilon^2$
and the excitonic Bohr radius for the length scale $a_x = 
\hbar^2\epsilon/\mu e^2$, where $\mu = m_{oe}m_{oh}/(m_{oe}+m_{oh})$. 
The reduced mass $\mu$ is determined by optical masses where
$3/m_{o\kappa} = 2/m_{\kappa}^\perp +1/m_{\kappa}^\parallel$.

The density functional theory \cite{Hohenberg64} expresses 
the ground state energy in terms of electron and hole densities 
only and when applied to the Hamiltonian (\ref{eqn:modham}) it gives: 
\begin{eqnarray}
E[n_{\lambda}({\bf r})] &=& T[n_{\lambda}({\bf r})]+
             E_{xc}[n_{\lambda}({\bf r})] \nonumber\\
&&+\frac{1}{2}\sum_{\lambda,\lambda'}
q_\lambda q_{\lambda'}\int d{\bf r}d{\bf r'}
\frac{n_{\lambda}({\bf r})n_{\lambda'}({\bf r}')}{|{\bf r}-{\bf r'}|} \ ,
\label{eqn:energy}
\end{eqnarray} 
$n_\lambda({\bf r})$ is the density of the component $\lambda$ 
[see Eq.~(\ref{eqn:modham})]. The Coulomb energy is split 
into a direct  Hartree term and an exchange-correlation 
term. The exchange-correlation energy is considered in the local density 
approximation which can be expressed as follows: 
\begin{equation}  
E_{xc}[n_{\lambda}({\bf r})] \approx \int d{\bf r}\;\epsilon_{xc}
   [n_{\lambda}({\bf r})] \sum_\lambda n_{\lambda}({\bf r})  \ .
\end{equation}
$\epsilon_{xc}$ is obtained from the ground state energy calculations 
of a uniform neutral plasma with {\em equal} electron and hole densities.
In our case charge neutrality can be locally broken due to
different electron and hole masses. However, because of a strong Coulomb 
interaction the two densities differ only slightly and, therefore, it is 
reasonable to calculate $\epsilon_{xc}$ substituting an averaged pair density  
\begin{equation}
   \label{eqn:density}
   n_{e\mbox{-}h}(r) = \frac{1}{2}\sum_{\nu,\sigma}
   \Bigl[ n_{e,\nu,\sigma} ({\bf r})
   +n_{h,\nu,\sigma}({\bf r})\Bigr] .
\end{equation}
Vashishta and Kalia \cite{Vashishta82} 
have shown that the exchange-correlation 
energy of a homogeneous electron-hole liquid is nearly 
independent of several band characteristics of semiconductors, 
such as valley degeneracy, effect of anisotropy, and electron-hole 
mass ratio. The reason is that the anisotropic 
contribution from the exchange energy cancels out the
contribution of the correlation energy for these band
characteristics. Vashishta and Kalia further fitted $\epsilon_{xc}$ 
from their self-consistent calculations to a simple analytic expression
\begin{equation}
   \label{eqn:epsxc}
   \epsilon_{xc}(r_s) = \frac{a + b r_s}{c+ dr_s +r_s^2} \ ,
\end{equation}
where $a=-4.8316$, $b=-5.0879$, $c=0.0152$, $d=3.0426$, and
$r_s = (3/4\pi a_x^3 n_{e\mbox{-}h})^{\frac{1}{3}}$ is
the dimensionless distance between carriers.

The densities in Eq.~(\ref{eqn:energy}) are related to the solution 
of the Kohn-Sham equations \cite{Kohn65} which are self-consistent 
one-particle Schr\"odinger equations:
\begin{eqnarray}
\label{eqn:K-S}
&&\left[ -\mu\hat{K}_\lambda + V_\lambda(n_\lambda,{\bf r})\right]
\psi_{\lambda i}({\bf r})=\epsilon_{\lambda i} \psi_{\lambda i} 
({\bf r})\ , \\
&&V_\lambda = q_\lambda\int d{\bf r'}\;\frac{\sum_{\lambda'} 
  q_{\lambda'} n_{\lambda'}({\bf r'})}
{|{\bf r}-{\bf r'}|}+ \frac{1}{2}
\phi_{xc}[n_{e\mbox{-}h}({\bf r})] \ , \nonumber \\
&&\phi_{xc}[n] = \frac{d(n\epsilon_{xc}[n])}{d n}\ ,\nonumber \\
&& n_{\lambda}({\bf r}) =\sum_{i=1}^{N_\lambda}
|\psi_{\lambda i}({\bf r})|^2 \ .
\nonumber 
\end{eqnarray}
The sum of the eigenvalues $\epsilon_{\lambda i}$ over all occupied 
states $(\lambda,i)$ does not give the energy of the molecule but 
it can be directly related to it by:
\begin{eqnarray}
\label{eqn:K-Senergy}
E & = & \sum_{\lambda,i}\epsilon_{\lambda i} - \frac{1}{2} 
\sum_{\lambda,\lambda'}q_\lambda q_{\lambda'}\int d{\bf r}d{\bf r'} 
\frac{n_{\lambda}({\bf r})n_{\lambda'}({\bf r}')}{|{\bf r}-{\bf r'}|} +
\nonumber\\
&& \mbox{}+ \int d{\bf r}\; \Bigl\{\epsilon_{xc}[n_{e\mbox{-}h}({\bf r})] -
  \phi_{xc}[n_{e\mbox{-}h}({\bf r})]\Bigr\} n_{e\mbox{-}h}({\bf r}) \ . 
\end{eqnarray}
An approximate solutions to the Kohn-Sham equations can be obtained 
by using a variational ansatz for $\psi_{\lambda i}$ and minimizing
the energy (\ref{eqn:K-Senergy}). Alternatively, one can choose 
to solve the Kohn-Sham equations directly imposing a self-consistency 
requirement. We calculate a few simple cases in both ways and then 
resort to a variational approach in more complicated situations.

\section{Binding energy calculation for multiexciton molecules}

\subsection{Isotropic and equal electron and hole masses}

\begin{figure}[t]
\unitlength1cm
\epsfxsize=8cm
\begin{picture}(2,3)
\put(0.1,-3){\epsffile{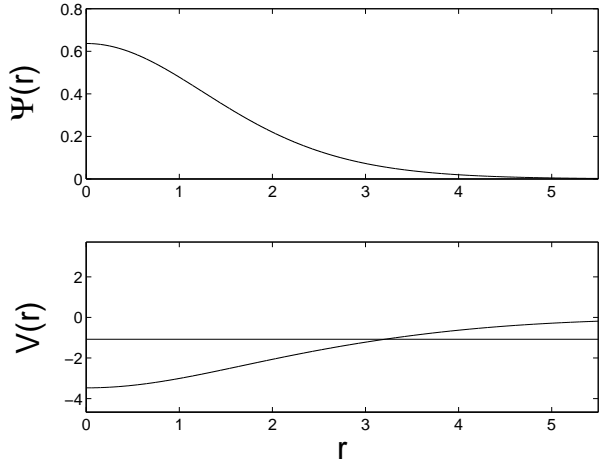}}
\end{picture}
\vspace{3cm}
\caption{The self-consistent numerical 1$s$ solution of the 
  Kohn-Sham equation for a molecule formed of six electron-hole pairs.
  The resulting effective potential is presented on the bottom panel 
  and the lowest eigenvalue is shown by a horizontal line.}
\label{fig:K-S}
\end{figure}

For isotropic masses ($m_{\kappa}^\parallel=m_{\kappa}^\perp\equiv 
m_\kappa$), we use the spherical symmetry of the ground state wave
function and the effective potential $V({\bf r})$:
\begin{equation}
\psi_{\lambda,nlm}({\bf r})=R_{\lambda, nl}(r)Y_{lm}(\theta,\phi) \ .
\end{equation}
Defining $\sigma_{e\mbox{-}h}=m_e/m_h$ and $R_{\lambda, nl}(r)=
\chi_{\lambda, nl}(r)/r$, we obtain the $l=0$ Kohn-Sham equations 
for electrons and holes:
\begin{eqnarray}
\left[-\frac{1}{1+\sigma_{e\mbox{-}h}} \frac{d^2}{d r^2} + 
V_e\right]\chi_{e,n0}(r) & = & \epsilon_{e,n0} \chi_{e,n0}(r)\ , 
\nonumber \\
\label{eqn:K-Seq}
\left[-\frac{\sigma_{e\mbox{-}h}}{1+\sigma_{e\mbox{-}h}} 
\frac{d^2}{d r^2} + V_h\right] \chi_{h,n0}(r) &=&
\epsilon_{h,n0} \chi_{h,n0}(r) \ .
\end{eqnarray}
The two equations differ only in the kinetic energy term because 
of unequal electron and hole masses. In this subsection we consider 
the case of equal masses $\sigma_{e\mbox{-}h} =1$. Then, we have only 
one Kohn-Sham equation to solve. This equation gives the same density 
profiles for electrons and holes and, therefore, the direct Hartree 
term vanishes identically. The self-consistent solution of 
the Kohn-Sham equation for the lowest eigenvalue is shown in 
Fig.~\ref{fig:K-S} for a molecule formed of 6 electron-hole pairs.

For the variational solution of the Kohn-Sham equations we take
the trial functions to be of `Fermi-Dirac' type:
\begin{eqnarray}
\mbox{1$s$ states:}\quad R_{1s}(r) &=& \frac{n_1}
{1+\exp(\frac{r-r_1}{\rho_1})}  \ , \nonumber \\
\mbox{2$s$ states:}\quad R_{2s}(r) &=& 
\frac{n_2(1-br)}{1+\exp(\frac{r-r_2}{\rho_2})}  \ .
\end{eqnarray}
Here, $r_1$, $\rho_1$, $r_2$, $\rho_2$ are variational parameters and 
$n_1,n_2$ are normalization constants. The choice of the 1$s$ 
wave-function is specially suited for large exciton molecules, 
which are described by a `droplet' model: the density is constant up 
to $r_1$ and then drops to zero in a surface layer of width $\rho_1$.
The results of the two methods for different molecule sizes are 
summarized in table \ref{tab1}.

\begin{figure}[t]
\unitlength1cm
\epsfxsize=8cm
\begin{picture}(2,3)
\put(0.1,-3){\epsffile{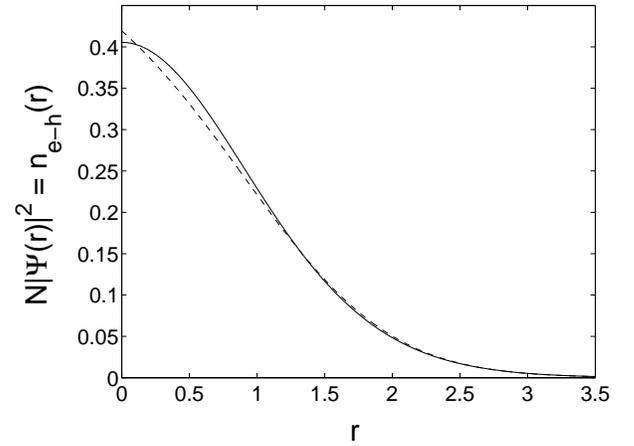}}
\end{picture}
\vspace{3cm}
\caption{The electron-hole pair densities profiles for numerical 
(solid line) and variational (dashed line) solutions 
  of the Kohn-Sham equations. 
  The results are for the most stable 6-exciton molecule.}
\label{fig:densities}
\end{figure}

The variational and exact solutions of the Kohn-Sham equation are in 
a very good agreement. The difference in energies does not
exceed 0.5\% for all molecule sizes. This agreement is not only
achieved in the binding energies but also in the wave-functions. 
The comparison of the electron-hole pair densities is plotted in 
Fig.~\ref{fig:densities} for a molecule formed of six electron-hole 
pairs. For small $r$ the variational trial functions have a linear 
behavior which explains the small difference at the center of the 
molecule between the exact solution and the variational solution. 
The third column of Table \ref{tab1} is a measure for the stability 
of the molecule against dissociation into the next smaller molecule 
and one free exciton. We see that only the 5- and the 6-pairs 
molecules are stable against this dissociation. 
The 6 electron-hole pairs molecule is in a state where all single 
particle states of the lowest shell are filled up. This represents 
the most stable configuration. The energy gain of 24\% of $E_x$
per one $e$-$h$ pair in a 6-exciton molecule is much higher
than a binding energy of an ordinary bi-exciton molecule for 
nondegenerate bands, which is only 1.7\% of $E_x$ 
\cite{Adamowski71,BRB}. If we go to larger molecules shell effects 
appear. The extra energy cost which is needed to put a further electron 
into the 2$s$ shell favors clearly the 6 electron-hole molecule. 
Shell effects appear also if the radii of different molecules are 
compared. Surprisingly, the radius gets smaller when we are filling up 
the 1s shell. This feature was also found by W\"unsche and co-workers 
\cite{Wunsche78b}. The radius of the molecule shows a sharp increase 
if a further electron-hole pair in the 2$s$ shell is added.

The calculated binding energies have to be compared to the ground 
state energy of the metallic electron-hole liquid. The ground state 
energy $E_{\rm g.s.}(r_s)$ was calculated in the RPA approximation 
in Ref.~\cite{Mike99b} with $E_{\rm g.s.}^{\rm min} = -1.55E_x$
at $r_s=0.92$. The ansatz (\ref{eqn:epsxc}) gives a close value
$E_{\rm g.s.}^{\rm min} = -1.6E_x$, which is reached at the density 
$r_s\approx 1.0$. Both these values are lower than the maximum gain 
from formation of a 6-exciton molecule which indicates that in 
the chosen approximation there is no intermediate phase of a dilute 
Bose gas of exciton molecules.

Although it is difficult to give an exact criterion for the 
applicability of the density functional approach to multiexciton 
molecules, it is expected that the theory works best at high densities, 
i.e.\ for  large molecules. Further, these densities have to be 
compared to the density for which the ground state energy per pair 
for an electron-hole liquid is minimal $r_s=0.92$. We expect that in 
the limit of large $N$, the density at the center of the molecule 
approaches this value. For a finite number of electron-hole pairs 
the density at the center will be higher than $r_s\approx 0.9$, 
since we must include a surface tension which increases the 
density. To see if such a picture of an electron-hole droplet is 
correct, we split the binding energy $E_B$ into a bulk and a 
surface term:
\begin{equation}
E_B = E_{bulk} + E_{surf} \ .
\end{equation}
We approximate the density by a uniform spherical density of radius
$r_M$. Then the bulk energy can be determined from the Fig.~2
of Ref.~\cite{Mike99b}. The surface energy is proportional to 
$r_M^2$:
\begin{eqnarray}
E_{bulk} &=&  N E_{\rm g.s.}[r_s(r_M)] \ , \nonumber \\
E_{surf} &=& 4\,\pi r_M^2 S \ .
\end{eqnarray}
Table \ref{tab2} presents the surface tension $S$ evaluated for 
different molecules. If an electron-hole droplet consideration 
of an exciton molecule is correct, then $S$ should be a constant. 
From table \ref{tab2} we can see that this is true for molecules 
formed of 5--7 electron-hole pairs but it fails for smaller molecules.

\subsection{Effect of electron-to-hole mass ratio}

\begin{figure}[t]
\unitlength1cm
\epsfxsize=8cm
\begin{picture}(2,3)
\put(0.1,-3){\epsffile{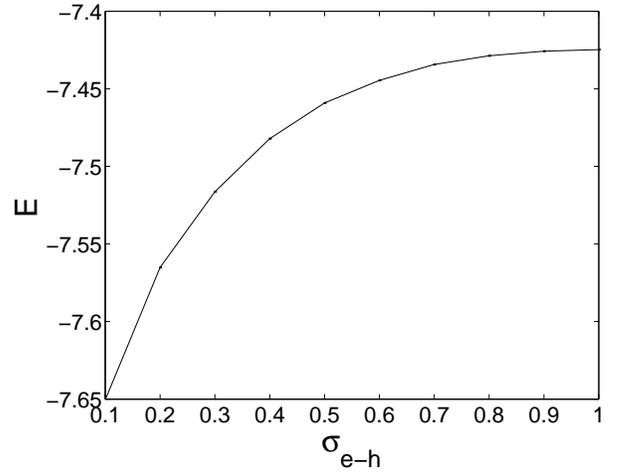}}
\end{picture}
\vspace{3cm}
\caption{The binding energy of a 6-exciton molecule versus 
the electron-to-hole  mass  ratio $\sigma_{e\mbox{-}h}$.}
\label{fig:sigma}
\end{figure}

In this subsection we lift the approximation of equal electron and 
hole masses but continue to replace the effective mass tensors by 
isotropic optical masses. The numerical procedure of solution of 
the coupled K-S equations (\ref{eqn:K-Seq}) was found to be unstable. 
In view of the results of the previous section, we expect, however, 
to obtain accurate results using again `Fermi-Dirac'-type variational 
wave-functions, but this time with different adjustable parameters 
for electron and holes. Our calculations are limited to the most stable 
molecule formed of six electron-hole pairs. We have varied 
$\sigma_{e\mbox{-}h}$ from $1$ to $0.1$. The results are plotted 
in Fig.~\ref{fig:sigma} and further details are presented shown in 
table III. The total energy of the 6-exciton molecule is lowered 
only by an amount of $0.23\,E_x$ or $4\,\%$ if $\sigma_{e\mbox{-}h}$ 
varies from $1$ to $0.1$. For CaB$_6$ the ratio of electron and hole 
optical masses is $\sigma_{e\mbox{-}h}=0.89$. For this value of 
$\sigma_{e\mbox{-}h}$ the electron and hole densities will be only 
slightly different and there is nearly no change in the binding energy
$E_B$. 

We can check validity of the obtained results by comparing them to the 
exact analytic criteria which have been derived in Ref.~\cite{Adamowski71}
for the problem of a bi-exciton molecule. Frequently these criteria 
are not satisfied by variational solutions. To check them we evaluate 
the energy functional (\ref{eqn:energy}) in the ground state and take 
it to be a function of $\sigma_{e\mbox{-}h}$: $E(\sigma_{e\mbox{-}h})$. 
From the form of the Hamiltonian (\ref{eqn:modham}) it can be shown 
\cite{Adamowski71} that $[\partial E(\sigma_{e\mbox{-}h})/\partial
\sigma_{e\mbox{-}h}]_{\sigma_{e\mbox{-}h} = 1} = 0$ and that 
$E(\sigma_{e\mbox{-}h})$ has to be a concave and monotonically
increasing function of $\sigma_{e\mbox{-}h}$. We see in 
Fig.~\ref{fig:sigma} that these two criteria are satisfied by our 
solution, which further supports validity of the obtained results.

\subsection{Effect of mass anisotropy}

\label{section:aniso}
According to the results of the previous subsection the binding energy 
changes only very little for unequal electron and hole masses. Therefore, 
for simplicity we assume that electron and hole dispersion  are described 
by the {\em same} anisotropic mass tensors. Otherwise, one has to deal 
with too many variational parameters. As in the previous section we limit 
our calculations to the 6-exciton molecule. We define the reduced 
masses $1/m^\perp=1/m_e^{\perp}+1/m_h^{\perp}$ and $1/m^\parallel = 
1/m_e^{\parallel}+1/m_h^{\parallel}$ and the ratio between the components 
of the reduced mass tensor $\sigma_a=m^\perp/m^\parallel$, with 
$m^\perp<m^\parallel$ or $\sigma_a<1$.

The variational trial functions are taken to be again of 
`Fermi-Dirac'-type, but now we use functions with a cylindrical 
symmetry. In different valleys the wave-functions are different. 
For the first valley they are defined as
\begin{equation}
\psi_{\kappa,1}({\bf r})=\frac{n_0}
    {1+\exp\left(\sqrt{\frac{y^2+z^2}{d_\perp^2}  
    + \frac{x^2}{d_\parallel^2}}-\rho\right)} \ .
\end{equation} 
$n_0$ is the normalization constant and $d_\perp$, $d_\parallel$ and 
$\rho$ are the three variational parameters. The wave-functions for 
particles in the other valleys are obtained by cyclical permutations 
of $x$, $y$ and $z$ in analogy with the kinetic energy operator 
$\hat{K}_\lambda$. This implies that the total density distribution 
is not spherically symmetric but is a superposition of three 
ellipsoidal distributions along the coordinate axes. 

The results of the minimization of the energy functional
(\ref{eqn:energy}) are presented in table \ref{tab4} for
$0.2<\sigma_a<1$. Varying $\sigma_a$ in this range produces 
an increase in the binding energy of $0.51\,E_x$ or $6\%$. 
For CaB$_6$ the mass ratio is $0.255$, which corresponds to an 
energy of $7.81\;E_x$ and thus to an increase in binding by 5\%.
Thus, mass anisotropy has a somewhat stronger effect on binding
energies of multiexciton molecules in the hexaborides than 
electron-to-hole mass ratio, though both effects 
give only small corrections 
to the simplest model with equal and isotropic masses.

\section{Exciton complexes bound to impurities}

It is well known that excitons in semiconductors 
are attracted to charged donor or acceptor 
impurities \cite{Kirczenow77,Dean76,Insepov77,Wunsche78b}. We 
investigate here the case of a monovalent donor impurity, which models 
the effect of La$^{3+}$-substitution in divalent-metal hexaborides 
CaB$_6$ and SrB$_6$. We add an external Coulomb potential to the 
Hamiltonian (\ref{eqn:modham}): 
\begin{equation}
\hat{V} = \sum_\lambda\int d{\bf r}\Psi^\dag_\lambda({\bf r}) 
  V_\lambda\Psi_\lambda({\bf r}) \ , \quad 
\hat{V}_\lambda=\frac{e q_\lambda}{\epsilon r}
\end{equation}
and assume equal and isotropic electron and hole masses. (In this case 
there is an obvious symmetry between donors and acceptors.)
At low impurity concentrations the electrons and holes are completely
localized at a single impurity. The impurity adds a heavy center to the 
molecule. Therefore we choose the variational wave-functions 
in the hydrogenic form:
\begin{eqnarray} 
R_e({\bf r}) &=& \frac{1}{\sqrt{\pi r_e^3}}\,
\exp\left(-\frac{r}{r_e}\right) \ , \nonumber \\
R_h({\bf r}) &=& \frac{1}{\sqrt{3\pi r_h^3}}\,
\frac{r}{r_h}\exp\left(-\frac{r}{r_h}\right),
\label{hole_shell}
\end{eqnarray}
with electrons and holes occupying $s$-wave and $p$-wave
orbitals, respectively, near a positively charged donor.
The results for the binding energies of multiexciton complexes 
are presented in table V. The second column shows 
the total energy of a complex $-E_B$, which includes one extra 
electron in addition to $N$ electron-hole pairs. All complexes 
with $N>1$ are stable against dissociation into a next smaller 
complex and a free exciton, since $|E_B(N+1)-E_B(N)|>E_x$. 
To compare the energy gain from a formation of a multiexciton
complex to the energy gain in a dense electron-hole plasma we 
need to subtract from $E_B$ the energy of a donor with a single 
electron, which is approximately $2E_x$. The third column shows 
that excitons gain the most of energy in a 5-exciton complex, 
which is again consistent with a filled shell argument.
The energy gain per one exciton in a 5-exciton complex exceeds
the energy gain in a 6-exciton molecule. Hence, upon decreasing 
the band gap charged impurities will work as nucleation centers
for electron-hole droplets and localized excitons will appear before 
condensation of multiexciton molecules in the bulk of a semiconductor.

The maximum possible energy gain per exciton in the 5-exciton complex 
$\Delta E=1.38E_x$ is still below the energy gain 
in a dense electron-hole 
plasma $|E_{\rm g.s.}^{\rm min}| = 1.55E_x$. Effective mass calculations 
have also been done for divalent donors (acceptors) with 4 excitons
being attracted, but we did not find any additional energy lowering for 
them. This result means that the direct first-order transition occurs 
under pressure without an intermediate excitonic phase. However, 
the two numbers are now closer to each other. In such a case so-called 
central cell corrections to the effective mass approximation, which 
have been estimated in Ref.~\cite{Mike99b} as $\Delta E_{\rm c.c.}
\sim 0.2E_x$ for a single exciton, can be sufficient to increase the 
energy gain per one exciton in a donor complex compared to the energy 
density in a bulk $e$-$h$ plasma. As a result, an intermediate state of 
localized excitons may appear under pressure. These localized exciton 
complexes bound to charged impurities carry uncompensated spin-1/2,
as suggested by the shell scenario. It is an interesting problem 
to understand whether such an intermediate phase can be responsible 
for the unusual relaxation effects in the hexaborides \cite{Gavilano}. 

Usually, in lightly doped semiconductors (e.g.\ P in Si) localized 
donor electrons develop {\it antiferromagnetic} correlations 
between nearest-neighbor donors.
This can be understood in the following way. For two 
donor atoms, which appear to be neighbors, the electronic 
hydrogen-type orbitals hybridize forming a lower bonding orbital 
and an upper antibonding orbital, similar to a hydrogen molecule. The 
two extra electrons will go to the nondegenerate bonding orbital and 
the Pauli principle requires them to have opposite spins.
We suggest that the opposite sign {\it ferromagnetic}
correlations will develop in a localized excitonic phase
of doped hexaborides. Their origin is in additional
multivalley degeneracy of electrons and holes in the hexaborides.
The bonding orbitals for two nearest-neighbor donors have 
six-fold total degeneracy. We expect that in such a case only 
4 excitons will be attracted to such a double center complex 
in order to fill the bonding electron orbitals.
The total spin of a double-impurity complex (0 or 1) comes
from a partially filled hole shell (two `holes' in the hole shell).
In the case of degenerate orbitals the Hund's rule plays the major
role and produces parallel alignment of spins of the two `holes.' 
Similar scenario applies also to divalent donors or acceptors.
(In the case of the hexaborides, Ca- or Sr-vacancies can play
a role of divalent acceptors.) Localized moments produced
in this way could be a source of the unusually fast NMR-relaxation
observed experimentally \cite{Gavilano} and lead to significant
sample-to-sample variations of ferromagnetic
moments in Ca$_{1-x}$La$_x$B$_6$ \cite{Terashima}. 

Estimation of the Hund's splitting in multiexciton complexes
is an interesting open question. Note, that ordinary semiconductors,
like Si or Ge, also have a multivalley structure of the 
conduction band. 
The degeneracy of donor orbitals in these cases is partially lifted by 
anisotropic central cell corrections, which select a nondegenerate 
lowest level and, hence, suppress ferromagnetic correlations.
This effect might be present in multiexciton complexes as well.
However, the difference between the two cases is in the shells 
which produce the total spin. In the multiexciton complexes 
bound to a donor, uncompensated spin is 
formed in the $p$-wave hole shell. Wave-functions for these 
states vanish at the origin Eq.~(\ref{hole_shell}) and have 
a small probability in the central cell. Therefore, the states in
the hole shell are better described in the effective mass theory 
than states in the electron shell and their splitting must be less 
significant.

\section{Conclusions}

We have shown that the multivalley degeneracy which is present
in both the conduction and the valence bands of the hexaboride 
materials leads to a number of peculiar effects:
(i) 6-exciton molecules are stabilized due to the shell
effect. Their binding energies are comparable to the exciton
binding energy in contrast to a weak binding energy of
a bi-exciton molecule. The energetics of the multiexciton complexes
is rather insensitive to the details of the band structure: 
electron-to-hole mass ratio and anisotropy in the effective mass tensors.
(ii) Multiexciton complexes are attracted to charged impurities.
They have a larger energy gain per one electron-hole pair than
a 6-exciton molecule. Therefore, upon decreasing a semiconducting 
gap excitons will first condense at donors and acceptors
producing an intermediate phase of localized excitons.
Though, the energy of this phase lies somewhat higher than the energy
gain in a dense electron-hole plasma, localized excitons
can be further stabilized by central cell effects or appear as 
a metastable phase at the first-order transition between
a semiconducting and a semimetallic state.
(iii) Uncompensated spins on multiexciton complexes show a tendency
towards ferromagnetic correlations, which is again a multivalley
effect. 

Our theoretical results indicate that a semiconducting phase
of divalent hexaborides must have a number of interesting
physical properties. Further experiments at ambient and 
applied pressure can shed more light on their relevance 
to the physics of hexaborides.

Financial support for this work was provided by Swiss National Fund.

\begin{table}
\caption{The binding energies $E_B$ of molecules consisting
 of $N$ excitons for isotropic and equal $e$ and $h$ masses.
 The equilibrium radius for a given molecule was estimated
 by $r_M =\int d{\bf r} r n({\bf r})/\int d{\bf r}n({\bf r})$.  
 All values are given in excitonic units.}
\label{tab1}
\begin{tabular}{cccccccccc} 
 Method & $N$ & $-E_B$ & $-\frac{1}{N}E_B$ & $r_M$ & $r_s(0)$&
 $\overline{r}_1$&
 $\overline{\rho}_1$&
 $\overline{r}_2$&
 $\overline{\rho}_2$\\ 
\tableline
   & 7 & 7.90 &  1.13 & 2.10 & 0.80& 1.2&0.79&2.1&1.19\\
 Variational  & 6 & 7.42 &  1.24 & 1.65& 0.83&1.4&0.72&-&-\\
   & 5 & 5.77 &  1.15 & 1.73  & 0.91&1.4&0.77&-&-\\
   & 4 & 4.24 &  1.06 & 1.78 & 1.00&1.4&0.80&-&-\\
   & 3 & 2.86 &  0.95 & 1,87  & 1.04&1.4&0.86&-&-\\
\tableline
        & 7 & 7.92 &  1.13 & 2.05  & 0.83\\
 Self-  & 6 & 7.43 &  1.24 & 1.64 & 0.84\\
consistent  & 5 & 5.78 &  1.16 & 1.70 & 0.93\\
        & 4 & 4.25 &  1.06 & 1.77  & 1.04\\
        & 3 & 2.86 &  0.95 & 1.87 & 1.10\\
\end{tabular}
\end{table}

\begin{table}
\caption{The surface tension for exciton molecules of different
     sizes.}
\label{tab2}
\begin{tabular}{crrrrr}
     $N$ & 7 &  6 & 5 & 4& 3 \\ 
     \tableline
     $4\pi S$ & 0.75& 0.79& 0.77&0.66&0.49 \\
\end{tabular}
\end{table}

\begin{table}
\label{tab3}
\caption{The binding energy of a 6-exciton molecule for different 
  values of the electron-hole mass ratio $\sigma_{e\mbox{-}h}$.
  $\overline{r}_e,\,\overline{r}_h,\,
  \overline{\lambda}_e,\,\overline{\lambda}_h$ are the  
  optimal values of the variational parameters.}
\begin{tabular}{ccccccc} 
 $\sigma_{e\mbox{-}h}$ & $-E_B$ & $-E_B/6$ & $\overline{r}_e$ &
 $\overline{\lambda}_e$&
 $\overline{r}_e$&
 $\overline{\lambda}_h$\\ 
\tableline
  1.0 & 7.42 &  1.24  & 1.40 & 0.72 & 1.40 & 0.72\\
  0.9 & 7.43 &  1.24  & 1.30 & 0.75 & 1.34 & 0.73\\
  0.8 & 7.43 &  1.24 & 1.32 & 0.75 & 1.36 & 0.72\\
  0.7 & 7.43 &  1.24 & 1.28 & 0.79 & 1.34 & 0.71\\
  0.6 & 7.45 &  1.24 & 1.30 & 0.77 & 1.30 & 0.70\\
  0.5 & 7.46 &  1.24 & 1.28 & 0.76 & 1.36 & 0.69\\
  0.4 & 7.48 &  1.25 & 1.28 & 0.76 & 1.38 & 0.67\\
  0.3 & 7.52 &  1.25 & 1.28 & 0.76 & 1.38 & 0.65\\
  0.2 & 7.57 &  1.26 & 1.30 & 0.74 & 1.38 & 0.62\\
  0.1 & 7.65 &  1.28 & 1.28 & 0.74 & 1.42 & 0.58\\
\end{tabular}
\end{table}

\begin{table}
\label{tab4}
\caption{The binding energy of a 6-exciton molecule for different
  values of the anisotropy parameter $\sigma_a$.
  $\overline{d}_\parallel$,  $\overline{d}_\perp$ and
  $\overline{\rho}$ are the values of the variational parameters.}
\begin{tabular}{cccccc} 
  $\sigma_a$ & $-E_B$ & $-E_B/6$ & 
  $\overline{d}_\parallel$ & $\overline{d}_\perp$& $\overline{\rho}$\\ 
\tableline
        1.0 & 7.42 & 1.24  & 0.52 & 0.52 & 1.94\\
        0.9 & 7.43 & 1.24& 0.52 & 0.55 & 1.78\\
        0.8 & 7.44 &  1.24 & 0.49 & 0.56 & 1.78\\
        0.7 & 7.45 &  1.24 & 0.47 & 0.57 & 1.76\\
        0.6 & 7.49 &  1.25 & 0.43 & 0.58 & 1.76\\
        0.5 & 7.54 &  1.26 & 0.40 & 0.59 & 1.73\\
        0.4 & 7.61 &  1.27 & 0.35 & 0.59 & 1.73\\
        0.3 & 7.73 &  1.29 & 0.30 & 0.60 & 1.69\\
        0.255 & 7.81 & 1.30& 0.27 & 0.60 & 1.66\\
        0.2 & 7.93 &  1.32 & 0.23 & 0.61 & 1.62\\
\end{tabular}
\end{table}

\begin{table}
\label{tab5}
\caption{The binding energies for different multiexciton complexes.
  $\overline{r}_e$,  $\overline{r}_h$ are the values 
  of the variational parameters for which the
  energy is minimized.}
\begin{tabular}{ccccc} 
$N$ & $-E_B$ & $-(E_B+2)/N$ & $\overline{r}_e$ & $\overline{r}_h$ \\ 
\tableline
     5 & 8.91 & 1.38 & 0.90 & 0.62 \\
     4 & 7.21 & 1.30 & 0.89 & 0.64\\
     3 & 5.62 & 1.21 & 0.87 & 0.66  \\
     2 & 4.17 & 1.09 & 0.82 & 0.69 \\
     1 & 2.86 & 0.86 & 0.73 & 0.74 \\
\end{tabular}
\end{table}

\newpage

\end{document}